# Topological surface states revealed by the Zeeman effect in superconducting UTe$_2$

Zhen Zhu[1], Hans Christiansen[2], Yudi Huang[1], Kaiming Liu[1], Zheyu Wu[3], Shanta R. Saha[4], Johnpierre Paglione[4,5], Alexander G. Eaton[3], Andrej Cabala[6], Michal Vališka[6], Rafael M. Fernandes[1,7], Andreas Kreisel[2,8], Brian M. Andersen[2], and Vidya Madhavan[1,5]*

1. Department of Physics and Materials Research Laboratory, Grainger College of Engineering, University of Illinois at Urbana-Champaign, Urbana, IL, USA.
2. Niels Bohr Institute, University of Copenhagen, DK-2200 Copenhagen, Denmark.
3. Cavendish Laboratory, University of Cambridge, Cambridge CB3 0HE, United Kingdom.
4. Maryland Quantum Materials Center, Department of Physics, University of Maryland, College Park, MD, USA.
5. Canadian Institute for Advanced Research, Toronto, Ontario, Canada.
6. Charles University, Faculty of Mathematics and Physics, Department of Condensed Matter Physics, Ke Karlovu 5, Prague 2, 121 16, Czech Republic.
7. Anthony J. Leggett Institute for Condensed Matter Theory, Grainger College of Engineering, University of Illinois at Urbana-Champaign, Urbana, IL, USA.
8. Department of Physics and Astronomy, Uppsala University, Box 516, 751 20 Uppsala, Sweden.

* vm1@illinois.edu

## Abstract:

Intrinsic topological superconductors with protected boundary modes obeying non-Abelian statistics constitute a vanishingly small class of quantum materials. A defining spectroscopic signature of such phases is the presence of in-gap topological surface states (TSS). However, despite extensive theoretical proposals, their unambiguous experimental identification has remained elusive. Here we use vector magnetic-field scanning tunnelling microscopy to obtain direct spectroscopic evidence of TSS in the spin-triplet superconductor UTe$_2$. Atomic-scale spectroscopy reveals striking site-dependent superconductivity: Te sites host a large in-gap density of states that nearly fills the superconducting gap, whereas neighboring atomic sites remain gapped. Upon application of a magnetic field, the in-gap states on the Te sites are selectively suppressed, yielding a spatially homogeneous superconducting state with a markedly deeper gap relative to zero field. This site-selective gap evolution is in quantitative agreement with theoretical predictions for TSS in UTe$_2$ that possess dominant Te-orbital character. Spectral-function calculations incorporating the Zeeman coupling reproduce the observed magnetic-field response. Our results provide a spectroscopic fingerprint of the long-sought TSS in superconductors and establish UTe$_2$ as a compelling system for exploring intrinsic topological superconductivity.

**Main text:**

Odd-parity superconductors have emerged as promising platforms for realizing intrinsic topological superconductivity[1-3]. In systems with inversion symmetry, odd-parity pairing is tied to spin-triplet superconductivity, which under certain circumstances can host non-trivial topological states protected by time-reversal or crystalline symmetries. Among them are gapless surface states taking the form of Majorana cones or Majorana flat-bands, which are composed of neutral Bogoliubov quasiparticles that are their own anti-quasiparticles[4-7]. Such in-gap topological surface states (TSS) are robust against symmetry-preserving perturbations. When the protecting symmetries are broken, a gap opens in the TSS, providing a direct signature of the nontrivial topology of the bulk superconducting phase. Despite the promise of topological superconductivity, only a small number of superconductors are believed to host odd-parity pairing[8,9], making the experimental detection of TSS challenging.

This scarcity of intrinsic candidate topological superconductors has led to extensive efforts to create engineered platforms for the explicit realization of Majorana modes. Most approaches have relied on heterostructures that combine spin–orbit coupling, magnetism, and proximity-induced superconductivity[10-12]. These platforms typically require fine tuning of multiple parameters, as exemplified by hybrid superconductor–semiconductor nanowires[13,14] and topological insulator–superconductor heterostructures[15-17]. In parallel, iron-based superconductors, in which s-wave superconductivity coexists with normal-state TSS, have been proposed as a single-material platform for realizing Majorana modes via self-proximity effects[18,19]. However, across all three classes of systems discussed above, a magnetic field is necessary to induce these topological bound states and the bulk superconducting order parameter remains topologically trivial.

The one exception is atomic chains assembled on superconducting substrates, which have been reported to host zero-energy end states suggestive of topological superconductivity[20,21]. Yet in these one-dimensional structures, the primary evidence for non-trivial topology relies on the observation of bound states localized at the chain ends. Given that these systems are composed of magnetic atoms and are inherently finite in length, conclusively establishing that the bound states are topological, rather than trivial Yu–Shiba–Rusinov or finite-size hybridized states, remains challenging[22]. By contrast, in intrinsic topological superconductors, the nontrivial topology arises directly from the bulk pairing symmetry, without the need for proximity effects or heterostructure engineering. In this case, the bulk–boundary correspondence applies in its most direct form, and robust topological surface states in three-dimensional systems are expected to emerge naturally in zero magnetic field.

$UTe_2$, a strongly correlated superconductor with triplet pairing[23,24], is considered a promising candidate for realizing a topologically non-trivial superconducting phase. Topological superconductivity in strongly correlated systems is expected to arise from unconventional pairing interactions rather than a phonon-mediated mechanism, such as those driven by spin fluctuations[25,26]. These pairing interactions can be highly anisotropic, which in turn can lead to nodal structures in the superconducting gap[7]. Such systems are known as nodal topological superconductors (n-TS), as exemplified by the A phase of the triplet superfluid $^3$He[27,28]. The boundary states in spin-triplet pair condensates arise from the odd-parity nature of the superconducting parameter. These states are protected by time-reversal symmetry such that the nodal points of the surface states can emerge from time-reversal-invariant-momenta of the surface Brillouin zone. Breaking of time-reversal symmetry will then lift the nodal points of the surface states and alter the density of states (DOS). The orthorhombic crystal structure of UTe$_2$ belongs to the $D_{2h}$ point group, whose irreducible representations allow four symmetry-distinct triplet pairing states, $A_{1u}$, $B_{1u}$, $B_{2u}$, and $B_{3u}$, with the latter three exhibiting nodal gap structures. Recent thermal transport and specific heat measurements are consistent with the presence of nodes in its superconducting order parameter[29-31]. In n-TS such as the $B_{2u}$ and $B_{3u}$ pairing symmetries of UTe$_2$[6,32,33], additional imprints of the surface states emerge from the surface projected bulk nodes, leading to spectral features reminiscent of Fermi arc states in Weyl semimetals[34,35].

The bulk–boundary correspondence implies that the presence of TSS constitutes a definitive fingerprint of topological superconductivity, but accessing such states requires probes that are intrinsically sensitive to surface electronic properties. More importantly, establishing that non-trivial topology underlies the presence of in-gap quasiparticles remains extremely challenging. Previous spectroscopic measurements have revealed a substantial residual DOS within the superconducting gap of UTe$_2$[36,37]. However, the microscopic origin of these in-gap states has remained elusive, as they may arise from non-trivial effects such as TSS or trivial effects including pair-breaking disorder. In this work, we address this challenge by examining the magnetic field response of the site-resolved superconducting gap at the atomic scale. Using vector-magnetic-field scanning tunnelling microscopy and spectroscopy (STM/STS), we uncover a site-selective suppression of the sub-gap spectral weight that originates from Zeeman coupling of the TSS, thereby providing a smoking-gun signature of non-trivial topology in superconducting UTe$_2$.

The cleaved surface of a UTe$_2$ crystal grown by the molten salt flux[38] (MSF) method ($T_c$ = 2.1 K) is shown in Fig. 1a. The surface is characterized by a quasi-one-dimensional arrangement of Te atoms (henceforth called chains) along the crystallographic *a*-axis (i.e., the [100] direction). We denote the in-plane direction perpendicular to the chains and the out-of-plane direction as the *c**- and *b**- axes, respectively, reflecting their ~24° rotation with respect to the crystallographic *c*- and *b*-axes[39]. Figure 1b presents the height profile extracted along the dashed line in Fig. 1a, and the corresponding d*I*/d*V* line cut reveals a pronounced spatial modulation of the superconducting gap (Fig. 1c). Establishing the atomic registry in real space enables us to resolve the superconducting gap distribution on different atomic sites. The chains observed in the topography, identified either by their elevated apparent height[36] or by dominant spectral weight near the Fermi level[40], can arise from Te1 or Te2 atoms, whereas the interchain sites consistently include U atoms (see Supplementary Fig. 1 for details). Based on this assignment, we define three representative atomic sites (marked in Fig. 1a): 'Te', corresponding to Te atoms on the chains; 'b-Te', corresponding to positions between adjacent Te atoms along the chain direction on the surface, and 'U' corresponding to the interchain sites. Their spatial arrangement, revealed in the height profile (Fig. 1b), aligns with their spectroscopic contrast: U sites occupy the valleys and exhibit deep superconducting gaps, whereas Te and b-Te sites appear at higher elevations and show much shallower gaps with enhanced sub-gap spectral weight, consistent with previous reports[36]. This allows us to associate the peaks and valleys in the topography with different atomic sites and the site-dependent gaps, as shown in Figs. 1b and c.

To gain microscopic insights into the superconducting gap distribution of $UTe_2$, we obtained spatially resolved tunnelling spectra (i.e., d$I$/d$V$ map) over a 3 × 3 nm$^2$ field of view sampled on a 32 × 32 grid (Fig. 1d), achieving ~1 Å spatial resolution, which is sufficient to resolve atomic-scale variations of the gap. Figure 1e shows the superconducting gap averaged over all spectra acquired within the measured region, exhibiting a substantial residual DOS at the Fermi level and a gap magnitude of ±0.25 mV. We find that the STM-measured residual DOS within the superconducting gap in our MSF-grown samples ($T_c$ = 2.1 K) is similar to samples grown by chemical vapor transport (CVT) with a lower $T_c$ of 1.6K[36,37], as shown in the Supplementary Fig. 2. Conversely, transport measurements show a much smaller residual resistivity ratio in the MSF-grown crystals[41,42] than in CVT-grown[23]. This suggests that the residual DOS within the gap is not a consequence of pair-breaking disorder, but an intrinsic feature of the electronic structure of $UTe_2$. By identifying the Te, b-Te, and U sites from the peaks and valleys in the topography acquired simultaneously with the d$I$/d$V$ map (Supplementary Fig. 3), we extract and average the corresponding spectra to obtain representative superconducting gaps at each of the three sites (Fig. 1f). Similar to the linecut in Fig.1c, we find that the in-gap residual DOS is the largest at the Te site (i.e., extremely shallow gaps), smaller at the b-Te site, and the smallest at the U site (i.e., deepest gaps).

Given the triplet character of the pairing state, a natural candidate to explain the intrinsic and distinct in-gap spectral weight associated with different atomic species is a TSS. Indeed, the existence of TSS in $UTe_2$ with $B_{2u}$ and $B_{3u}$ triplet-pairing symmetry has been theoretically proposed by multiple groups[6,32,43-45]. Taking the $B_{2u}$ state as an example, we calculated the surface DOS for Te and U. Our spectral function calculations reveal that the TSS residing within the superconducting gap are dominated by Te-derived states (Fig. 1g). Thus, the resulting calculated DOS of Te exhibits a more suppressed gap than that of U (Fig. 1h), consistent with our experimental data.

To obtain further evidence for non-trivial topology in the superconducting phase of $UTe_2$,

we study the response of the in-gap states to magnetic fields. Owing to their helical nature, the TSS in UTe$_2$ are expected to exhibit a distinctive magnetic-field response that serves as a key spectroscopic fingerprint of their existence[46-48]. The spin degree of freedom couples to the magnetic field via the Zeeman effect, which breaks time-reversal symmetry and thus transforms the surface state from massless to massive by opening a gap at the crossing point. The Zeeman effect, governed by the energy scale of ~$g\mu_B B$ (where $\mu_B$ is the Bohr magneton and $g$ is the gyromagnetic ratio), typically manifests at low energies and requires ultrahigh energy resolution for spectroscopic detection, rendering such measurements experimentally challenging. To probe this subtle spectroscopic signature, we track the magnetic-field dependence of the superconducting gap in UTe$_2$ across different strengths and directions, aiming to confirm the topological nature of the in-gap states and further constrain the underlying pairing symmetry.

We first investigate the evolution of the spatially averaged superconducting gap under magnetic fields applied along the *a*-axis. Figure 2a displays the averaged gaps over all sites obtained from a series of maps acquired under the indicated magnetic fields, all measured over the same defect-free region with identical tunneling conditions and grid resolution (see Fig. 1d). As the field increases, the residual DOS at the Fermi level is rapidly suppressed, manifested by a deeper gap with sharper coherence peaks. The field-induced gap deepening saturates around 0.65 T and starts to weaken above 0.8 T. Figure 2b provides a clearer visualization of the field-dependent evolution of the superconducting gaps using the same dataset as in Fig. 2a and confirms that the gap size remains nearly unchanged with magnetic field.

To quantitatively capture the gap evolution, we define the 'gap depth' $\alpha_d$ as the average of $\alpha_d^+$ and $\alpha_d^-$, where $\alpha_d^\pm$ represents the difference in DOS between the coherence peak at positive (negative) bias and the gap minimum (see 0 T spectrum in Fig. 2a). Using this definition, we extract the gap depth under different magnetic fields and summarize the results in Fig. 2c. The gap depth increases with field, reaching a maximum of more than twice the zero-field value at 0.8 T, before gradually decreasing.

To clarify the origin of this unconventional gap response, we first assess and rule out possible contributions from local in-plane vortices. A magnetic field applied along the *a*-axis (B//*a*) is expected to align vortices in the same direction, with an estimated spacing of ~50 nm at 0.8 T, which would typically result in a spatial modulation of the superconducting gap depth of the order of this spacing. d*I*/d*V* linecuts taken perpendicular to the *a*-axis and spanning 80 nm (Fig. 2d) under the indicated magnetic fields are shown in Fig. 2e, while the corresponding average gaps extracted from the linecuts are shown in Fig. 2f. As previously observed in Fig. 1c, the superconducting gap exhibits nanoscale inhomogeneity at zero field. Under magnetic fields applied along both the B//*a* and B//-*a*, the gaps deepen across the entire field of view, accompanied by enhanced coherence peaks and without any detectable vortex-related features (such as an increase in the zero-bias DOS). The lack of detectable vortices is not surprising since vortices created by an in-plane field may occur far from the surface. Based on these data and our theory, a simple hypothesis is that the gap deepening induced by a magnetic field along the *a*-axis, in the absence of any periodic variation expected from vortex formation, originates from the magnetic-field induced gapping of the TSS in UTe$_2$. This hypothesis is substantiated by site-resolved spectra, as shown below.

By extracting and averaging the spectra measured at the Te, b-Te, and U sites from d*I*/d*V* maps taken under magnetic fields applied along the *a*-axis (see details in Supplementary Fig. 3), we resolve the site-dependent evolution of the superconducting gaps (Fig. 3a). Consistent with our theoretical calculations, which indicate that the TSS has primarily Te-orbital character, we observe the most pronounced response on the Te sites, where the spectra evolve from a shallow suppression to a deep, well-defined gap with sharp coherence peaks as the field increases. While the b−Te sites exhibit a comparable but attenuated gap evolution, the response is markedly weaker on the U sites. This phenomenology is consistently reproduced across different combinations of samples and tips (Supplementary Fig. 4). The gap inhomogeneity across different atomic sites diminishes with increasing field and becomes nearly uniform at high fields, indicative of a more homogeneous superconducting state. This trend is further supported by the extracted gap depths under different fields (Fig. 3b), which confirm not only that the field-induced gap enhancement is the strongest on Te sites, but also that all three sites converge to nearly the same gap depth in the range from 0.8 T to 0.95 T.

To visualize the evolution of the superconducting gap under magnetic field in real space, we extract and plot the gap depth for each spectrum in our maps as a function of field. Figure 3c displays the gap depth map at zero field, revealing a prominent modulation, with the minimum gap depth located at the Te sites and the maximum at the U sites. Upon applying a 0.65 T field along the *a*-axis, the overall gap depth increases and becomes markedly more uniform across the field of view (Fig. 3d). This is further supported by datasets at other fields presented in Supplementary Fig. 5. The corresponding difference map, obtained by subtracting the zero-field gap depth map from that at 0.65 T, reveals that the enhancement is primarily concentrated along the Te chains, particularly at the Te sites (Fig. 3e). A more detailed view of the local spectral evolution was obtained by acquiring high-resolution d*I*/d*V* line cuts along the same trajectory shown in Fig. 1a under varying fields (Fig. 3f). With increasing field, the DOS at the Fermi level on the Te chains is progressively suppressed, consistent with the observed enhancement of the gap depth.

Combining the above real-space imaging of superconducting gap variations, we identify the enhanced gap depth around Te sites as a direct Zeeman-induced response of the TSS in UTe$_2$. For a relativistic *g*-factor of 2, the Zeeman energy reaches approximately 0.1 meV at 0.8 T, which is comparable to the energy scale of the superconducting gap ($\Delta \approx 0.25$ meV). The fact that the most pronounced gap enhancement occurs at Te sites further supports the scenario that the TSS is dominated by Te orbital weights, in agreement with our theoretical calculations. Moreover, the gap size on the U sites primarily reflects the underlying bulk superconductivity of UTe$_2$, as they contribute minimally to the TSS and exhibit negligible variation with magnetic field. At high magnetic fields, the conductance spectra on Te and b−Te sites converge to that of the U sites, implying that only the surface states are gapped out by the field, while the bulk superconductivity remains unaffected (Fig. 3a).

The Zeeman effect is expected to manifest under magnetic fields applied along different directions. To explore this, we apply the magnetic field along the $c^*$-axis, which lies within the cleavage plane but is perpendicular to the Te chains (Fig. 1a). Under this configuration, a similar signature of Zeeman-induced gap evolution is observed even in the presence of an in-plane vortex. We obtain d$I$/d$V$ maps both with and without a magnetic field over the same measurement region (dashed box in Fig. 4a), which contains numerous defects concentrated near the center, likely serving as pinning centers for the in-plane vortices induced by the applied magnetic field. As shown in Supplementary Fig. 6a, the zero-bias d$I$/d$V$ map reveals DOS modulations at zero field. To isolate magnetic field effects from this intrinsic spatial variation, we average the spectra along the $c^*$-axis at each point of the map, yielding the integrated d$I$/d$V$ line cut shown in Fig. 4b, where a superconducting gap is visible throughout the entire mapped region. Upon applying a 0.8 T magnetic field along the $c^*$ direction, the integrated d$I$/d$V$ line cut (Fig. 4c, see Supplementary Fig. 6b for the zero bias d$I$/d$V$ map) reveals a zero-energy vortex bound state extending over ~40 nm along the $a$-axis, corresponding to a Ginzburg–Landau coherence length of $\xi \approx 20$ nm, in agreement with previous out-of-plane vortex results[49]. Because of the defect-induced vortex pinning, a similar in-plane vortex is observed in the same region under a 0.8 T field applied along the -$c^*$ direction (Fig. 4d).

We now compare spectra taken on Te sites at spatially identical locations (see details in Supplementary Fig. 7). Spectra obtained at the vortex core and on both sides of the core are analyzed separately. As shown in Fig. 4e, the averaged superconducting gap at the Te sites continues to deepen on both sides of the vortex under magnetic field (upper and lower panels), regardless of field direction, similar to the response observed for fields applied along the $a$-axis. Meanwhile, zero-energy bound states emerge at the vortex core (middle panel). The gap deepening under fields applied along the $c^*$-axis also originates from the Zeeman response of the TSS in UTe$_2$, but is weaker than the response under fields along the $a$-axis of the same magnitude. This behavior may originate from $g$-factor anisotropy inherent to systems with low crystalline symmetry, such as the orthorhombic UTe$_2$. As a result, the Zeeman coupling becomes direction dependent, leading to anisotropic responses in the quasiparticle spectrum. Moreover, because of the low-symmetry of the crystal, all the components of the d-vectors characterizing the $B_{2u}$ and $B_{3u}$ triplet states are non-zero, which ensures that the TSS responds to fields applied along any direction. As a final check we apply an out-of-plane field, which in our case corresponds to the *b*\*-axis. The Zeeman response of the TSS is also evident on one side of the vortex under an out-of-plane magnetic field (Supplementary Fig. 8), as reported in recent studies[49-51]. The spatial asymmetry in response to an out-of-plane field is intriguing but beyond the scope of this work.

The field-induced response of the superconducting gap is captured by introducing a Zeeman term into the spectral function calculations in a model containing electronic states originating from U atoms and Te atoms. Within this framework, we simulate the evolution of TSS for $B_{2u}$ pairing symmetry under a magnetic field applied along the *a*-axis (Figs. 5a-c). When time-reversal symmetry is broken by the Zeeman field, the Te-dominated TSS become gapped, reflecting the lifting of Kramers degeneracy and the resulting mass acquisition of the surface quasiparticles. In contrast, the U-derived band remains largely unaffected, consistent with its weaker coupling to the field and its more bulk-like character. With increasing magnetic field strength, the Zeeman-induced gap in the Te-derived surface states grows systematically, leading to a pronounced redistribution of low-energy spectral weight. The integrated DOS for Te and U, shown in Figs. 5d-f, captures the response of the superconducting gap under Zeeman coupling. Specifically, the Te-derived DOS develops a pronounced superconducting gap with increasing field, consistent with the experimental observations. A similar effect is obtained in simulations for $UTe_2$ with $B_{3u}$ pairing symmetry (Supplementary Fig. 9), indicating the robustness of Zeeman-driven responses for the in-gap TSS.

Our study provides direct spectroscopic evidence for the presence of TSS in $UTe_2$, revealing a field-induced deepening of the superconducting gap that is strongly site-dependent and most pronounced at the Te atom sites. This observation is consistent with theoretical predictions that the TSS possess dominant Te-orbital character in the triplet superconducting state. The gap response persists under different magnetic field orientations, highlighting the robustness of the surface state response. Complementary spectral function calculations incorporating Zeeman terms reproduce the experimental trends and further support the topological origin of the observed modulation. Taken together, these findings establish the Zeeman sensitivity of TSS in $UTe_2$ and demonstrate an effective approach for spectroscopically probing their topological nature. They also provide stringent constraints on the underlying pairing symmetry and advance the identification of bulk topological superconductivity in strongly correlated systems. Finally, they reveal a little-explored intertwining between topological states and the spatial inhomogeneity of the superconducting gap, which should be present in other multi-orbital triplet superconductors.

## Methods

Single crystals of $UTe_2$ grown by a molten flux method were used. The growth and characterization are mentioned in detail elsewhere[38]. The STM measurements were performed using a Unisoku STM system operating under ultra-high vacuum and equipped with a vector magnetic field ($H_x$:4T, $H_y$:1T, $H_z$:9T). The $UTe_2$ single crystals were cleaved in-situ at about 90 K and then transferred immediately into the STM head. All the measurements were carried out at 300 mK using tungsten tips treated with in situ heating. d$I$/d$V$ signals were acquired by a standard lock-in amplifier with modulation of 40 µV at 991 Hz.

**Acknowledgments:** STM studies at the University of Illinois, Urbana-Champaign were supported by the US Department of Energy, Office of Science, Office of Basic Energy Sciences, Materials Sciences and Engineering Division under award number DE-SC0022101. V.M. acknowledges partial support from Gordon and Betty More Foundation's EPiQS Initiative through grant GBMF4860 and the Quantum Materials Program at CIFAR where she is a Fellow. A. K. acknowledges support by the Danish National Committee for Research Infrastructure (NUFI) through the ESS-Lighthouse Q-MAT. H. C. acknowledges support from the Novo Nordisk Foundation Grant No. NNF20OC0060019. A.G.E. acknowledges support from the Henry Royce Institute for Advanced Materials through the Equipment Access Scheme enabling access to the Advanced Materials Characterization Suite at Cambridge, grant numbers EP/P024947/1, EP/M000524/1 & EP/R00661X/1; and from Sidney Sussex College (University of Cambridge). M.V and A.C. acknowledge support by the Czech Science Foundation GAČR under the Junior Star Grant No. 26-21795M (STiUS). Crystals were grown and characterized in MGML (mgml.eu), which is supported within the program of Czech Research Infrastructures (project No. LM2023065). Research at the University of Maryland was supported by the Gordon and Betty Moore Foundation's EPiQS Initiative Grant No. GBMF9071, the NIST Center for Neutron Research, and the Maryland Quantum Materials Center.

**Author contributions:** Z.Z. and V.M. conceived the experiments. Z.W., S.R.S., J.P., A.G.E., A.C. and M.V. provided the characterized single crystals. Z.Z., Y.H. and K.L. obtained the STM data. Z.Z., Y.H. and V.M. carried out the analysis and H.C., R.M.F., A.K., and B.M.A. provided the theoretical input on the interpretation of the data. Z.Z., and V.M. wrote the paper with input from all authors.

Competing interests: The authors declare no competing interest.

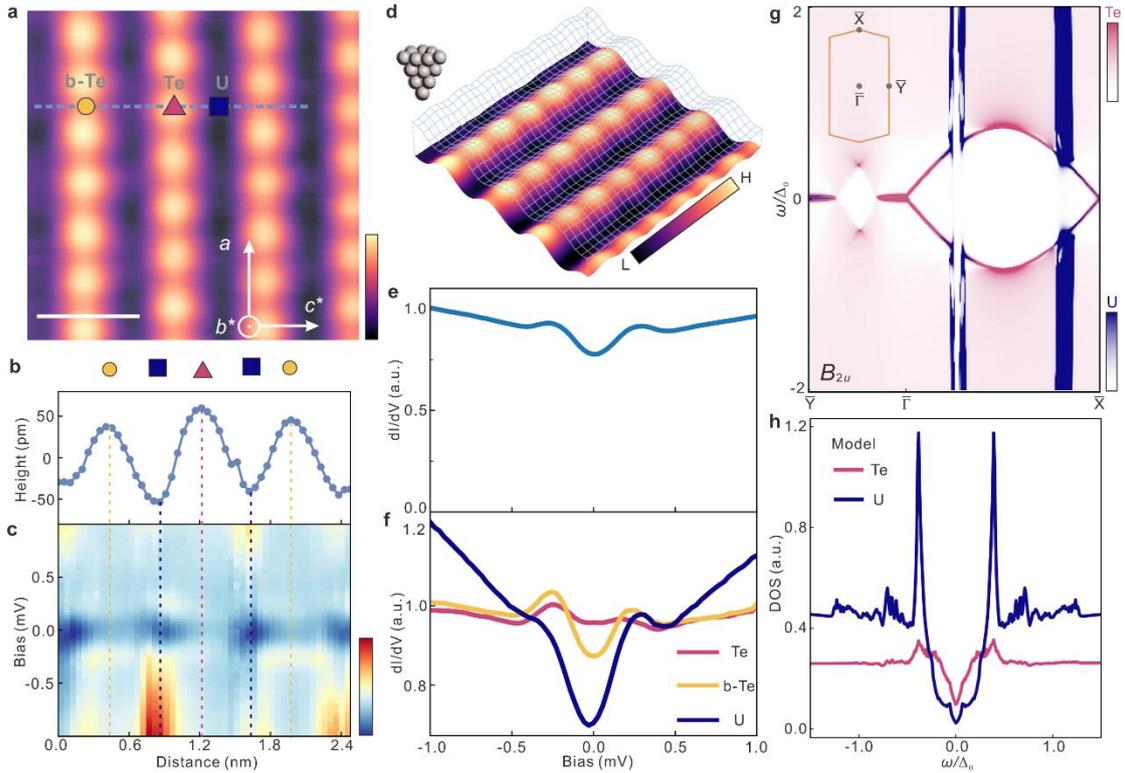

**Figure 1 | Spatially modulated superconducting gap and surface-state distribution of UTe$_2$. a**. Atomic-resolution images acquired on the surface of UTe$_2$. The one-dimensional chains formed by Te atoms are oriented along the crystallographic a-axis (V=10 mV, I=100 pA). The red triangles, yellow circles, and blue squares represent the three representative sites: Te, b-Te (between Te atoms), and U, respectively. Scale bar, 1nm. **b**. Height profile extracted along the dashed line in **a**. The peaks correspond to the Te and b-Te sites, while the valleys are associated with the U sites. **c**. dI/dV line cut taken along the same dashed line in **a**. A superconducting gap opens at the Fermi level, appearing deeper on U sites with lower DOS (darker blue) and shallower on Te sites with higher DOS (lighter blue). **d**. Schematic illustration of spatially resolved maps. A spatial map of the superconducting gap was obtained by measuring dI/dV spectra on a 32 × 32 grid over a 3 nm × 3 nm region (same as **a**). **e**. Gap spectrum obtained by averaging all spectra from the dI/dV map measured over the region shown in **d**. A large residual DOS is revealed at the Fermi energy, accompanied by coherence peaks located at ±0.25 mV (V=1 mV, I=100 pA). **f**. Site-resolved superconducting gaps on Te, b-Te, and U sites. Gap spectra, averaged over three representative site categories across the dI/dV map, reveal a progressive reduction of zero-bias DOS from Te (pink) to b−Te (yellow) to U (navy), reflecting the spatial inhomogeneity of the superconducting state (V=1 mV, I=100 pA). **g.** Calculated spectral functions for the $B_{2u}$ triplet phase of UTe$_2$. The topological surface states are dominated by Te weight. **h.** Simulated DOS versus energy for the superconducting $B_{2u}$ phase. The enhanced Te-derived DOS within the superconducting gap arises from surface states.

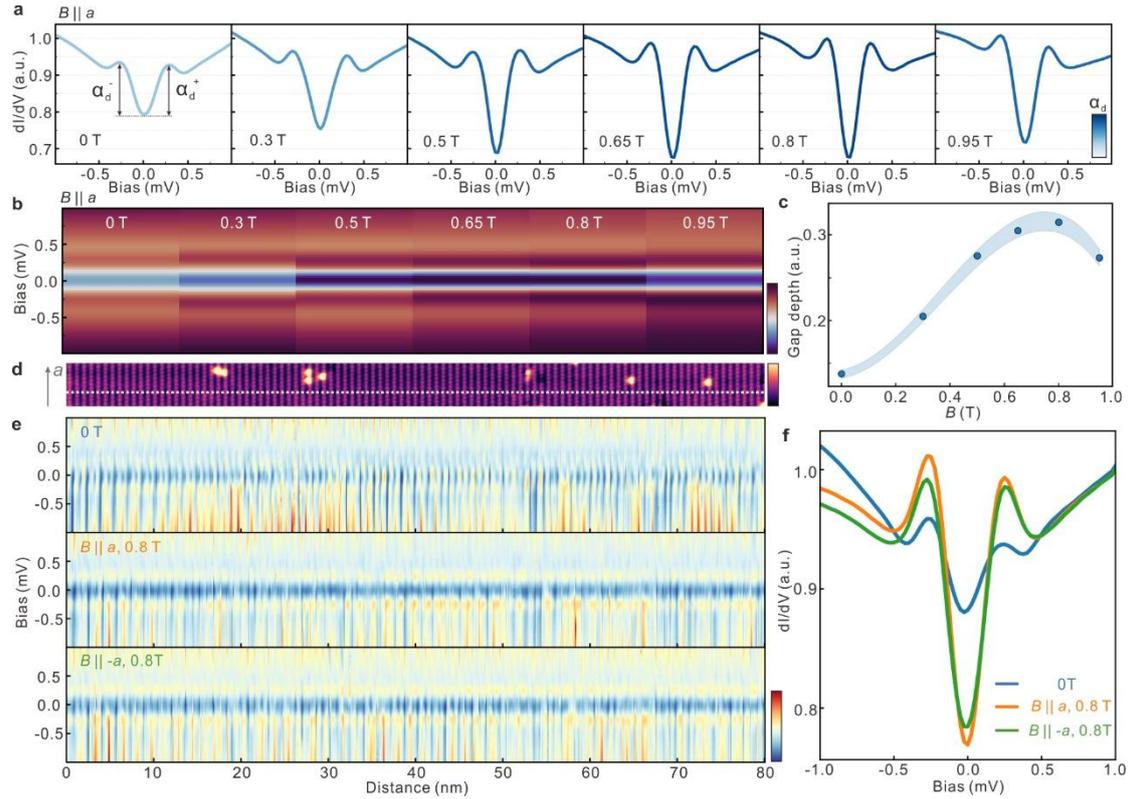

**Figure 2 | Evolution of the overall superconducting gap under magnetic fields along the *a*-axis. a.** Superconducting gaps acquired under the indicated magnetic fields. Each gap was obtained by averaging all spectra in the corresponding d*I*/d*V* map acquired from the same area (*V*=1 mV, *I*=100 pA), following the procedure shown in Fig. 1d. The superconducting gaps become deeper and the coherence peaks sharpen with increasing field, reaching saturation around 0.65 T and subsequently weakening above 0.8 T. The difference in DOS between the coherence peak at positive (negative) bias and the gap minimum is defined as $\alpha_d^+$ ($\alpha_d^-$), and the gap depth $\alpha_d$ is defined as the average: $(\alpha_d^+ + \alpha_d^-)/2$. **b.** Color map representation of the spectra shown in **a**. The gap size remains nearly unchanged with field, whereas the gap depth exhibits a pronounced field-induced enhancement. **c.** Magnetic-field dependence of the gap depth $\alpha_d$. The depth increases by up to 2.5 times its zero-field value under a magnetic field. **d.** Large-area topography (80 nm × 5 nm) of UTe$_2$ (*V*=40 mV, *I*=20 pA). **e.** d*I*/d*V* line cuts acquired along the same dashed line in **d** under the indicated magnetic fields. The superconducting gap displays spatial inhomogeneity between atomic sites at zero field. Upon applying 0.8 T along the ±*a* directions, the overall gap depth shows uniform enhancement across the large area, with no apparent in-plane vortices observed (*V*=1 mV, *I*=100 pA). **f.** Averaged spectra from line cuts in **e** under the corresponding fields. The spectra measured under 0.8 T along the *a* (orange) and the -*a* directions (green) exhibit nearly identical gap deepening compared to the zero-field spectrum (blue).

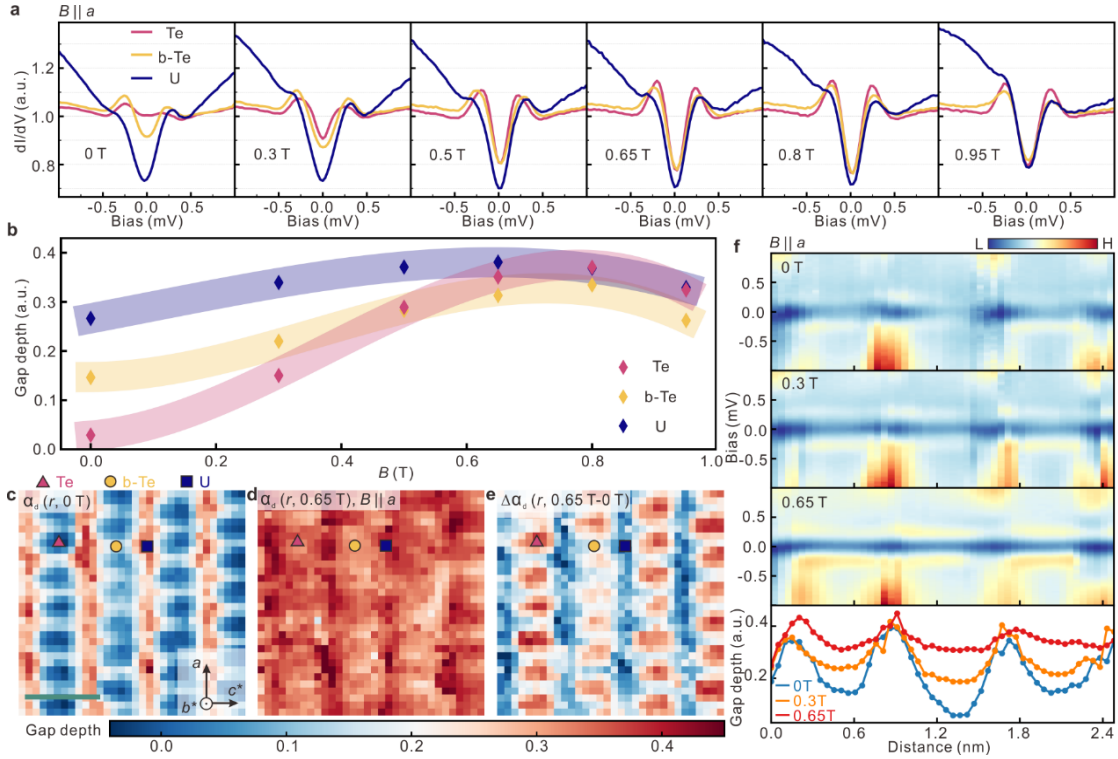

**Figure 3 | Field-induced gapping of topological surface states. a.** Site-resolved superconducting gaps acquired under the indicated magnetic fields. The spectra at each site were extracted from d$I$/d$V$ maps and subsequently averaged. The gap averaged at the Te sites deepens markedly with increasing field, evolves more gradually at the b-Te sites, and changes only slightly at the U sites, while at high fields all three gaps become nearly identical ($V$=1 mV, $I$=100 pA). **b.** Evolution of the averaged gap depth at different sites. The gap depth increases with field at all three locations, with a rate inversely correlated with their distance from the Te atoms. At high fields, the gap depth saturates and reaches similar values across all sites. **c.** Gap depth map at 0T. The gap depth exhibits a lattice-modulated variation in real space, with the minimum depth at the Te sites and the maximum at the U sites. Scale bar, 1 nm. **d.** Gap depth map at 0.65 T. The gap becomes uniformly enhanced across the area, in contrast to the spatial inhomogeneity observed at zero field. **e.** Difference map of the gap depth between 0.65 T and 0 T. The field-induced enhancement of the gap depth is most prominent on the Te sites, indicating that the topological surface states are gapped out under magnetic field. **f.** d$I$/d$V$ line cuts acquired along the dashed line in Fig. 1a under the indicated magnetic fields, together with the corresponding extracted gap depths. The spatial inhomogeneity of the superconducting gap weakens progressively with increasing field and eventually becomes uniform ($V$=1 mV, $I$=100 pA).

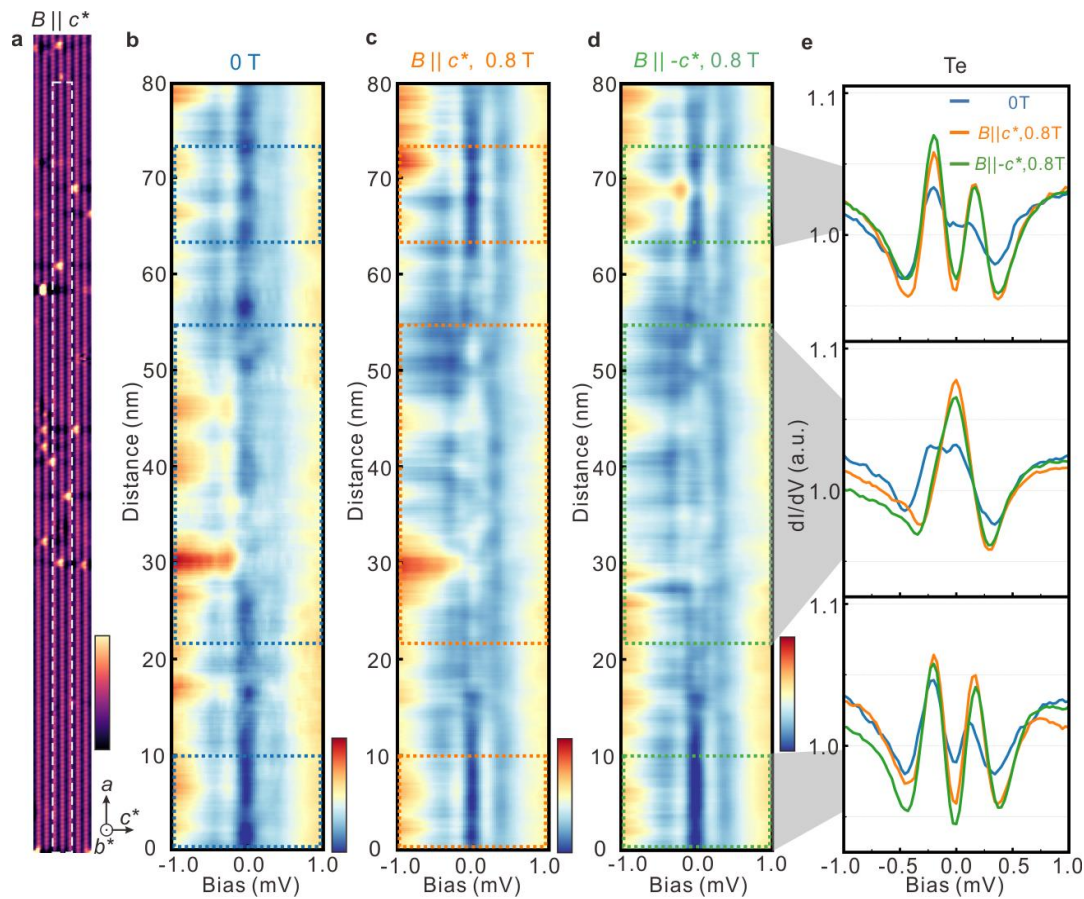

**Figure 4 | Anisotropic response of topological surface states under magnetic fields along the *c*\*-axis. a.** Large-area topography (6 nm × 85 nm) with numerous defects concentrated in the center (*V*=40 mV, *I*=20 pA). **b.** d*I*/d*V* line cut obtained by averaging the spectra along the *c*\*-axis from the d*I*/d*V* map acquired over the dashed-box region in **a** under zero field (*V*=1 mV, *I*=100 pA). A superconducting gap is visible along the entire line cut, showing suppression in the central region due to defect-bound states. **c, d,** Same as **b**, but acquired under 0.8 T applied along the *c*\* and -*c*\* directions, respectively. In-plane vortices appears at nearly the same location (central dashed boxes), each exhibiting a zero-energy vortex bound state (*V*=1 mV, *I*=100 pA). **e**. Averaged spectra of Te sites extracted from the d*I*/d*V* maps corresponding to the dashed-box regions marked in **b-d**, under the indicated magnetic fields (*V*=1 mV, *I*=100 pA). The superconducting gap deepens on both sides of the vortex under a magnetic field (upper and lower panels), while a pronounced zero-energy peak emerges at the vortex core (middle panel).

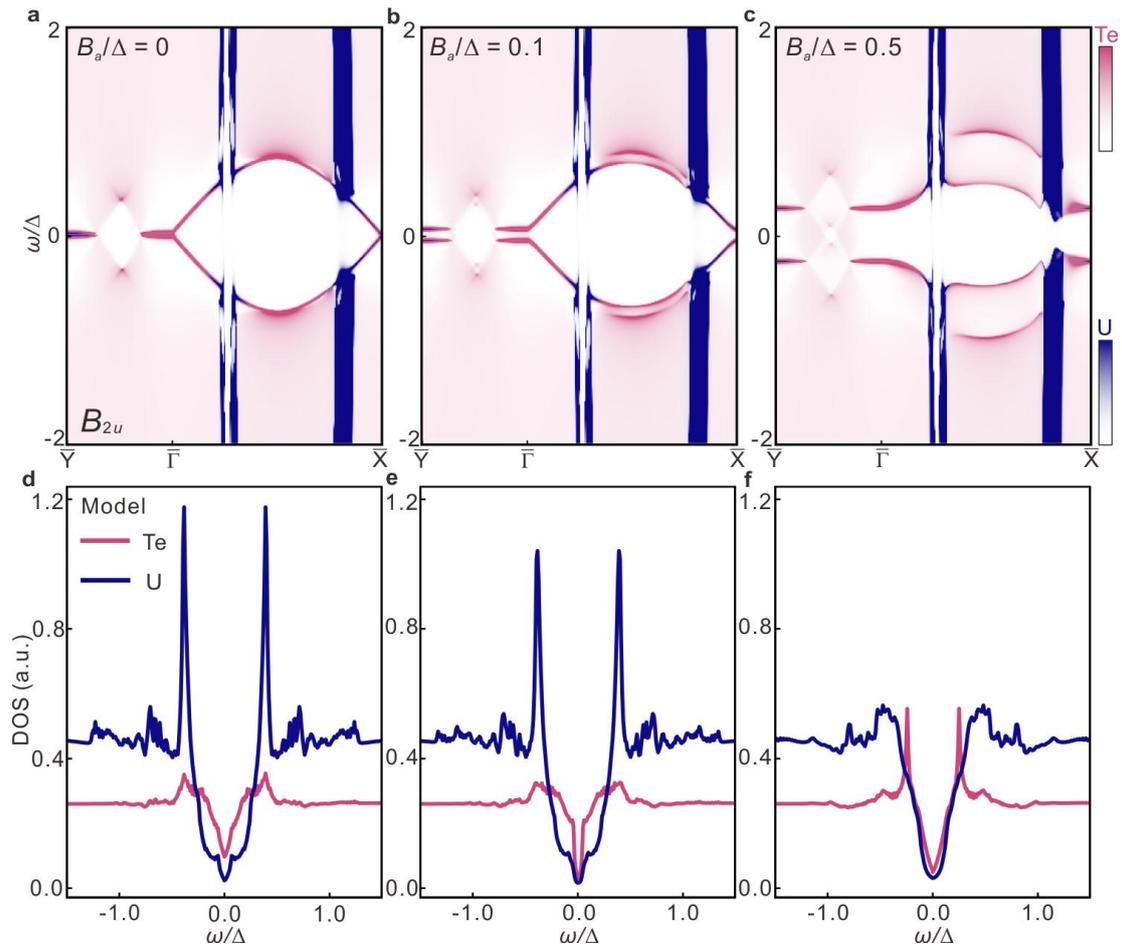

**Figure 5 | Zeeman effect on the topological surface states of UTe₂. a-c.** Spectral functions of Te and U in the $B_{2u}$ triplet superconducting phase with the indicated Zeeman energy. The topological surface states dominated by Te are gradually gapped out by the Zeeman effect. **d-f.** Integrated DOS versus energy corresponding to **a-c**. As the Zeeman energy increases, the Te-derived sub-gap DOS associated with the topological surface states becomes suppressed.

# Supplementary Information for

# Topological surface states revealed by the Zeeman effect in superconducting UTe$_2$

**Table of Contents**

**Supplementary Notes:**
Theoretical modelling of the topological surface states in UTe$_2$

**Supplementary Information Figures:**

**Supplementary Notes:**

**Theoretical modelling of the topological surface states**

The starting point of the theoretical modelling of the topological surface states and their response to Zeeman fields is a tight-binding model based on DFT calculations and consistent with recent quantum oscillation measurements[1]. This model describes the electronic structure near the Fermi energy in the bulk of the system. The superconducting triplet pairing with either $B_{2u}$ or $B_{3u}$ symmetry is captured by including lowest-order superconducting pairing terms consistent with the symmetries of the system and a quasi-particle spectrum exhibiting point nodes, the details of which are described in Ref. 2.

In our STM measurements, we cannot determine whether the quasi-one-dimensional chains observed in the topography originate from Te1 or Te2 atoms. Although DFT calculations indicate that the density of states near the Fermi level is dominated by Te2[1], the Wannier functions evaluated in the vacuum exhibit a larger amplitude above Te1 than above Te2[3]. Therefore, under small-bias tunneling conditions, the topographic contrast is more likely associated with Te1 sites, whereas the measured low-energy tunneling spectra predominantly probe the Te2-derived density of states at the Fermi level.

The (01-1) surface plane, spanned by the (100) and the (011) directions, is misaligned with the conventional unit cell of the crystal. The surface, therefore, presents a non-trivial cross-section of the unit cell. To properly capture the electronic properties on the surface, we therefore first transform to a unit cell that aligns with the surface plane. In this new coordinate system, we partition the bulk Hamiltonian in terms of a set of principal layers such that the first layer corresponds to the surface plane. We then use an iterative Green's function technique[4] to obtain the surface Green's functions (SGF)[2].

Assuming translation invariance on the surface, the SGF is diagonal in the momenta $\vec{k} = k_a m_a + k_{c^*} m_{c^*}$ where $m_a = (\frac{1}{a}, 0, 0)$ and $m_{c^*} = (0, \frac{1}{b}, \frac{1}{c})$. Note that since we perform a Fourier transform in 2D, there is an extra freedom in how we choose the out-of-plane component of the Brillouin zone (BZ). In particular, one can choose $m_a$ and $m_{c^*}$ in such a way that the BZ lies in the plane of the surface, without changing the corresponding SGF.

The SGF exhibit poles near the Fermi level at $\vec{k}$ close to the time-reversal invariant momenta (TRIM)[2], indicating the appearance of the TSS. These appear in pairs, related by time-reversal symmetry, and their dispersion close to the TRIM points are well described in terms of a massless Dirac Hamiltonian. For more details on the low-energy theory of the TSS, see the supplementary information from Ref. 5. The Zeeman field breaks time-reversal symmetry and therefore couples the two surface states. This coupling introduces a mass term in the Dirac Hamiltonian significantly decreasing the density of states at the Fermi Level. Since the TSS have weight mainly on the tellurium sites, this effect should be much stronger on the LDOS probing the tellurium atoms.

**Supplementary references:**

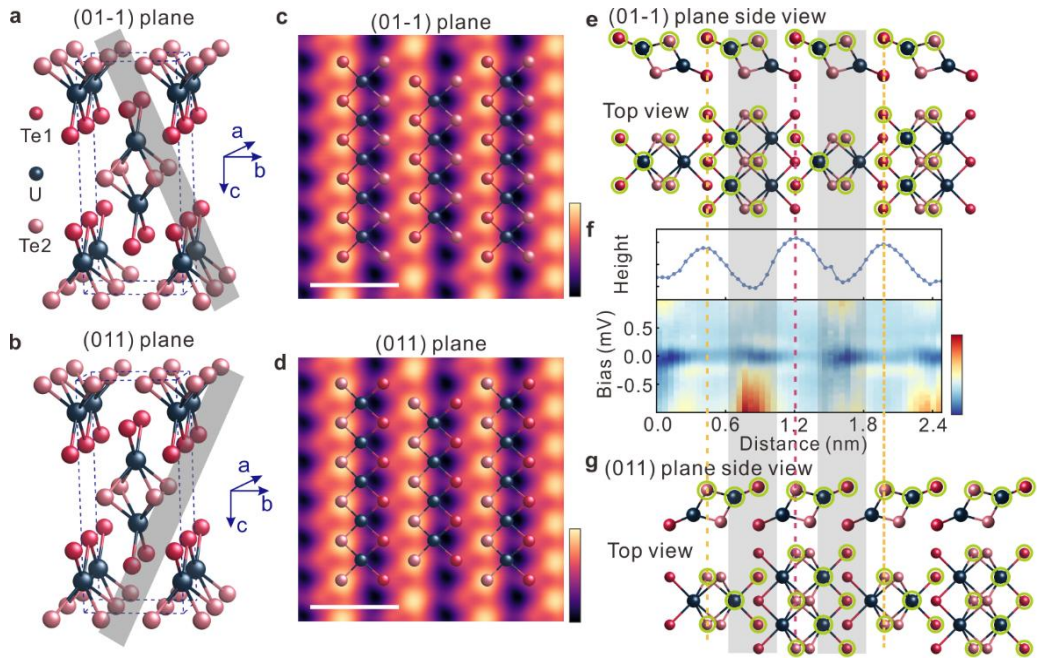

**Supplementary Figure 1 | Cleave plane assignment and site-dependent superconducting gap distribution in UTe$_2$. a, b,** Schematic crystal structure of UTe$_2$, with the (01-1) and (011) cleavage planes indicated by grey shading, respectively. Each U (black) atom is coordinated by two inequivalent Te sites, labeled Te1 (red) and Te2 (pink). **c,** Atomically resolved topographic image of UTe$_2$ showing both Te1 and Te2 chains ($V$ = 50 mV, $I$ = 100 pA). Based on the crystallographic structure, Te1 atoms lie slightly above Te2 atoms in the cleavage plane. If the brighter chains are attributed to Te1, this suggests a (01-1) cleavage. The topmost atomic model of the (01-1) surface is overlaid. Scale bar, 1nm. **d,** Same as **c**, but assuming the bright chains arise from Te2 atoms. Based on calculations showing that the density of states at the Fermi level is dominated by Te 2p orbitals[1], the brighter chains are attributed to Te2, suggesting a (011) cleavage. The topmost atomic structure of the (011) surface is overlaid. Scale bar, 1nm. **e,** Side and top views of (0-11) plane. Atoms highlighted with green circles indicate the topmost layer shown in **c**. **f,** Same as Fig.1b, c, showing the height profile and d$I$/d$V$ line cut along dashed line in Fig.1a. **g,** Side and top views of (011) plane. Atoms highlighted with green circles indicate the topmost layer shown in **d**. Comparing **e-g**, either Te1 or Te2 chains could correspond to the observed topographic peaks, so we label the atoms on the bright chains as Te sites, which exhibit shallow gaps. Regardless of whether the cleavage plane is (01-1) or (011), the valleys in the topography align with the positions between U atoms. These are therefore labeled as U sites, which exhibit deeper gaps.

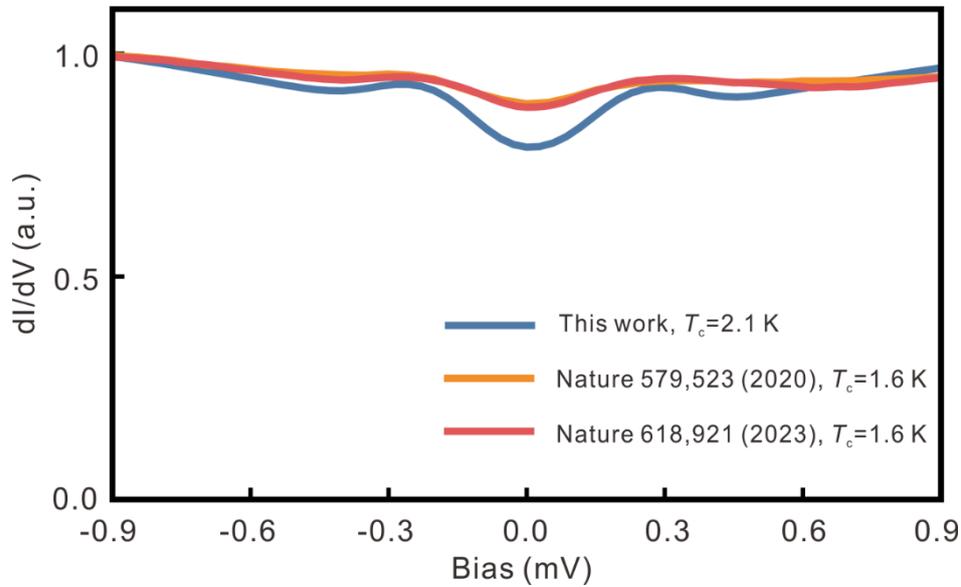

**Supplementary Figure 2 | Comparison of the superconducting gaps in UTe$_2$ crystals grown by molten salt flux (MSF) and chemical vapor transport (CVT).** The superconducting gap measured on the MSF-grown sample ($T_c$ = 2.1 K) in this work (blue line, same as Fig. 1e) exhibits a slightly deeper gap compared with previous studies on CVT-grown samples[6,7] ($T_c$ = 1.6 K, orange and red lines), with all measurements performed at 300 mK. Considering the enhanced $T_c$ and the much smaller residual resistivity ratio of the MSF-grown crystals[8], the large residual density of states observed within the gap is more likely to originate from the intrinsic electronic structure of UTe$_2$ rather than crystalline disorder.

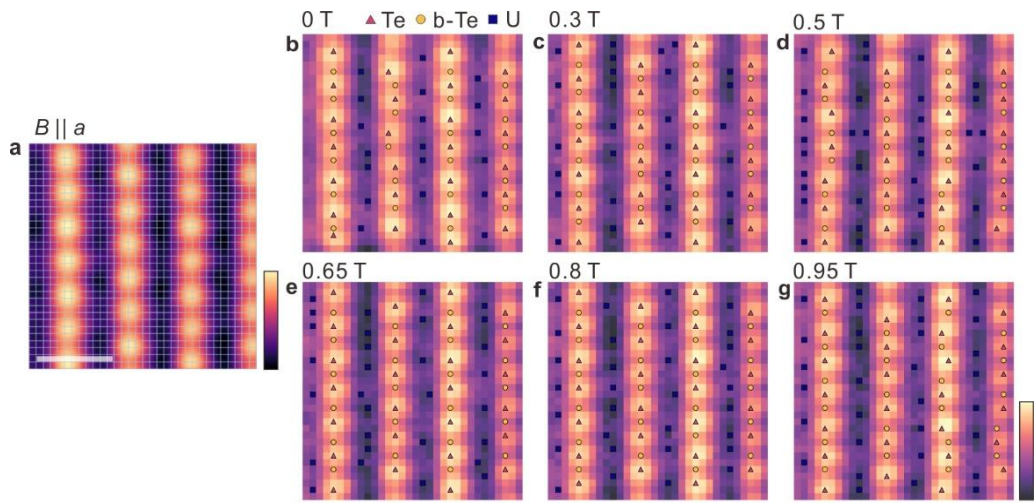

**Supplementary Figure 3 | Real-space identification of Te, b-Te, and U sites in d$I$/d$V$ maps under magnetic fields along the *a*-axis.** a. Topographic image of the area where the d$I$/d$V$ map was acquired. A 32 × 32 grid is overlaid, with a d$I$/d$V$ spectrum measured at each grid point ($V$ = 10 mV, $I$ = 100 pA). Scale bar, 1 nm. **b-g,** Topographies acquired simultaneously with each d$I$/d$V$ map under the indicated magnetic fields. Te sites (red triangles) and U sites (navy squares) are identified via two-dimensional peak and valley detection, respectively. b-Te sites (yellow circles) are defined as the valleys located between adjacent Te sites along the *a*-axis ($V$ = 1 mV, $I$ = 100 pA).

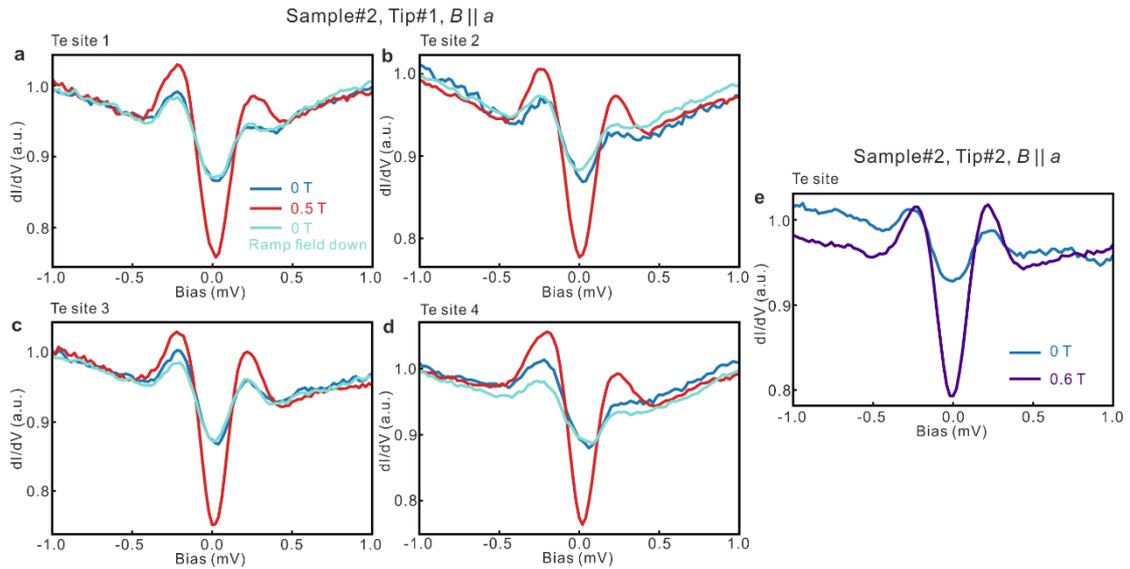

**Supplementary Figure 4 | Field-induced gap deepening in sample#2 with different tips under magnetic fields along the *a*-axis. a-d,** d*I*/d*V* spectra acquired on various Te sites using tip #1 under the indicated magnetic field. The superconducting gap deepens with increasing field, and recovers to zero-field gap when the field is ramped back down (*V* = 1 mV, *I* = 300 pA). **e,** d*I*/d*V* spectra acquired on Te site with tip #2 under the indicated magnetic fields. Similar field-induced gap deepening is observed (*V* = 1 mV, *I* = 300 pA).

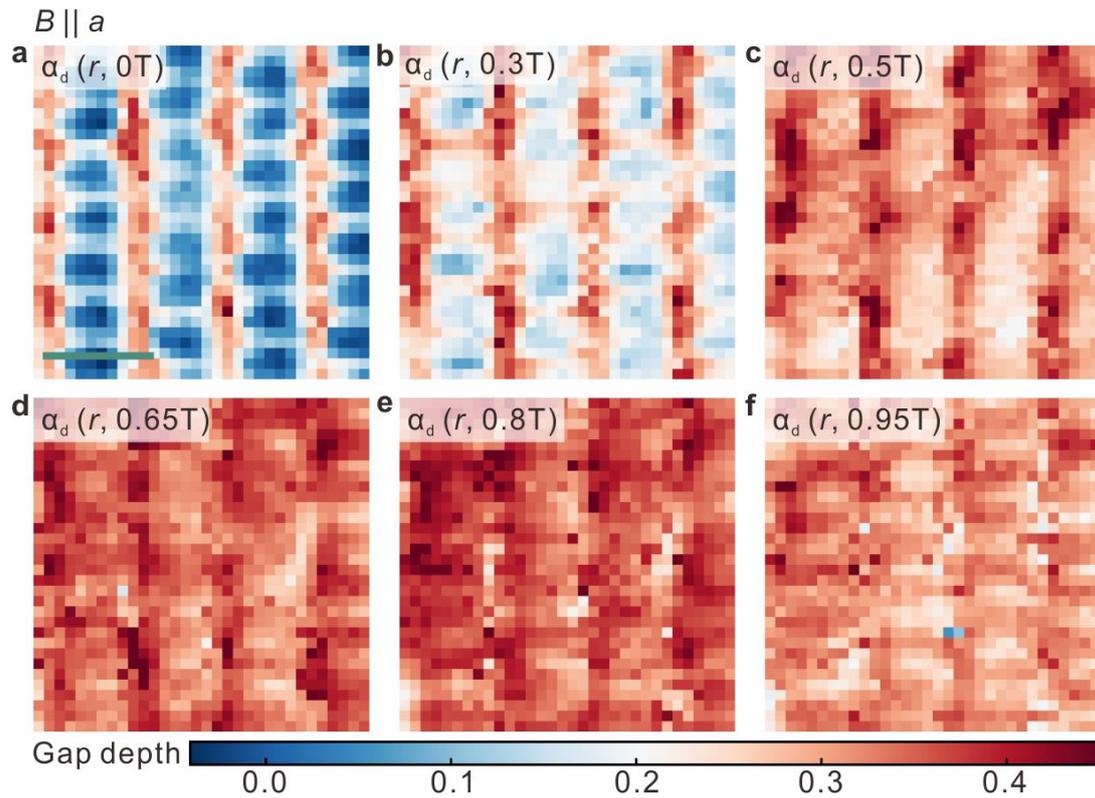

**Supplementary Figure 5 | Gap depth maps under magnetic fields along the *a*-axis. a-f,** Gap depth maps extracted from d*I*/d*V* maps under the indicated magnetic fields. As the field increases, the gap depth at Te sites progressively increases, and above 0.5 T, the gap depth becomes nearly uniform across all sites. Scale bar, 1 nm.

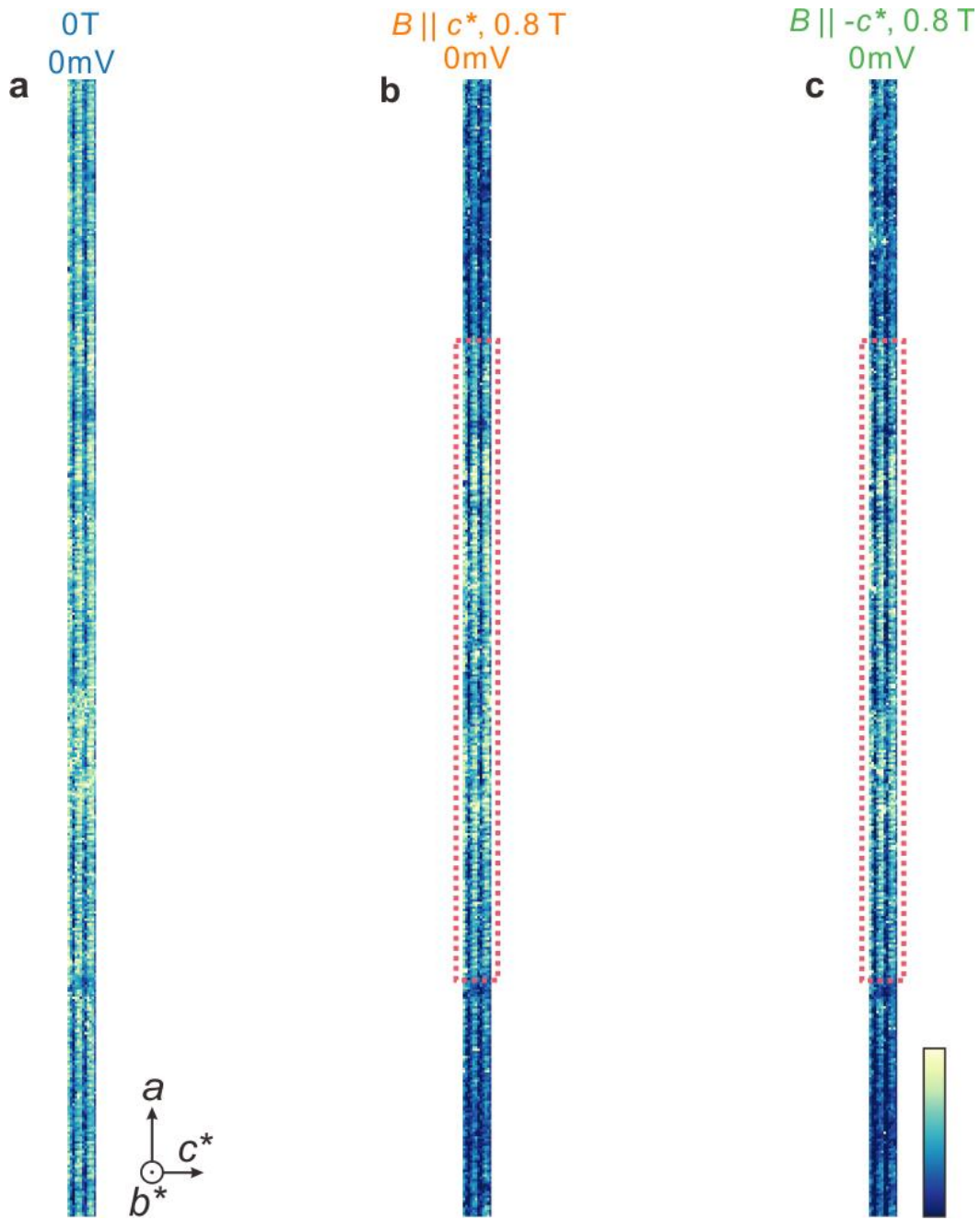

**Supplementary Figure 6 | Zero-bias d$I$/d$V$ maps acquired under magnetic fields along the $c^*$-axis. a.** Zero-bias d$I$/d$V$ map acquired over the dashed-box region in Fig. 4a under zero field. The DOS at the Fermi level is uniformly modulated by the Te chains ($V$=1 mV, $I$=100 pA). **b, c,** Same as **a**, but acquired under 0.8 T applied along the $c^*$ and -$c^*$ directions, respectively. In-plane vortices appear in the orange dashed boxes, exhibiting enhanced DOS under the indicated magnetic fields (V=1 mV, I=100 pA).

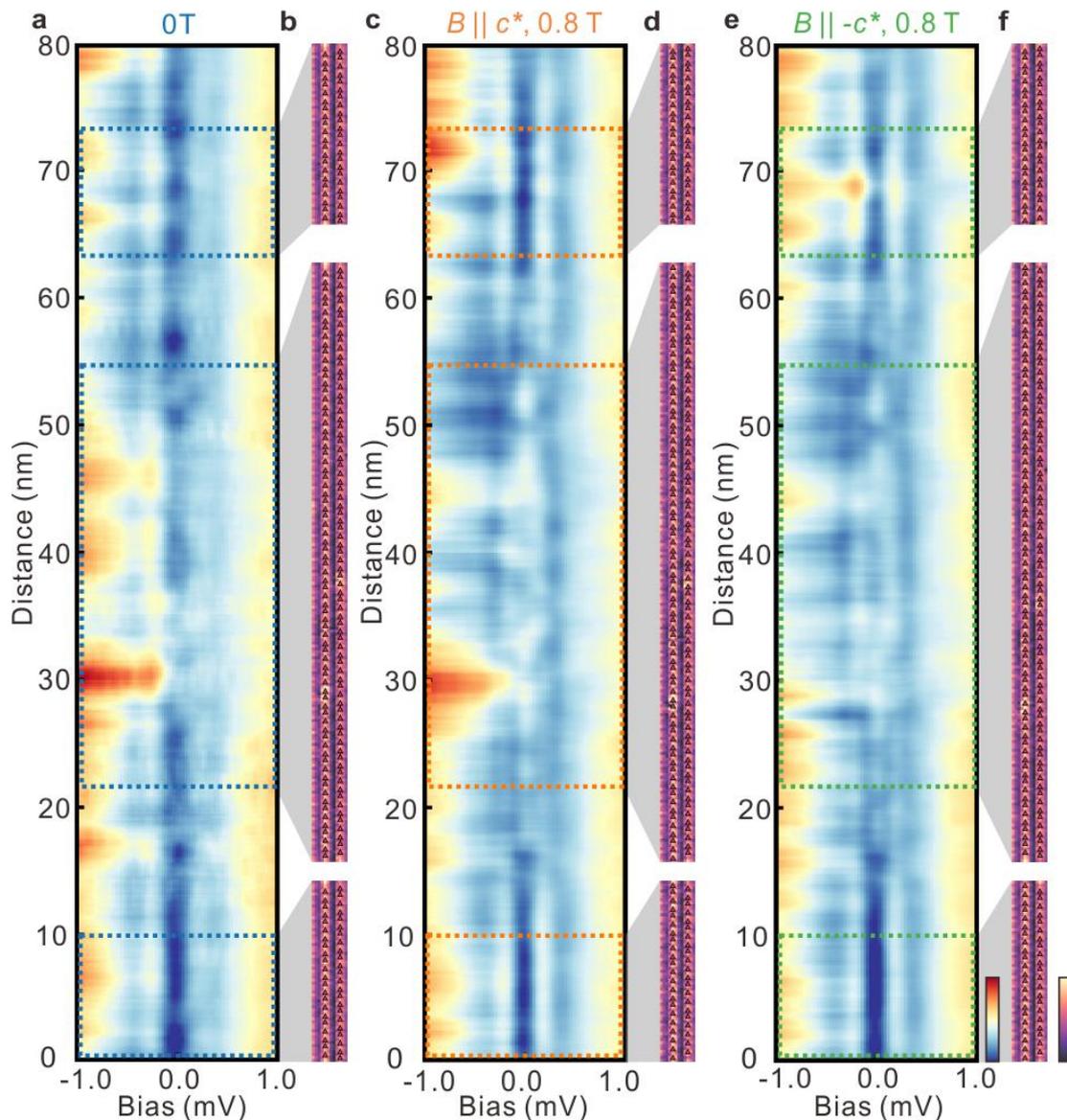

**Supplementary Figure 7| Real-space identification of Te sites in selected d$I$/d$V$ map regions under magnetic fields along the $c^*$-axis. a,** Same as Fig. 4b. **b,** Topographies acquired simultaneously with the zero-field d$I$/d$V$ map and extracted from the region indicated by the dashed boxes in **a**. **c,** Same as Fig. 4c. **d,** Same as **b**, but with 0.8T field applied along the $c^*$ direction and extract from the regions indicated by the dashed boxes in **c**. **e,** Same as Fig. 4d. **f,** Same as **b**, but with 0.8T field applied along -$c^*$ direction and extract from the regions indicated by the dashed boxes in **e**. Te sites (red triangles) in **b**, **d** and **f** are identified via two-dimensional peak detection.

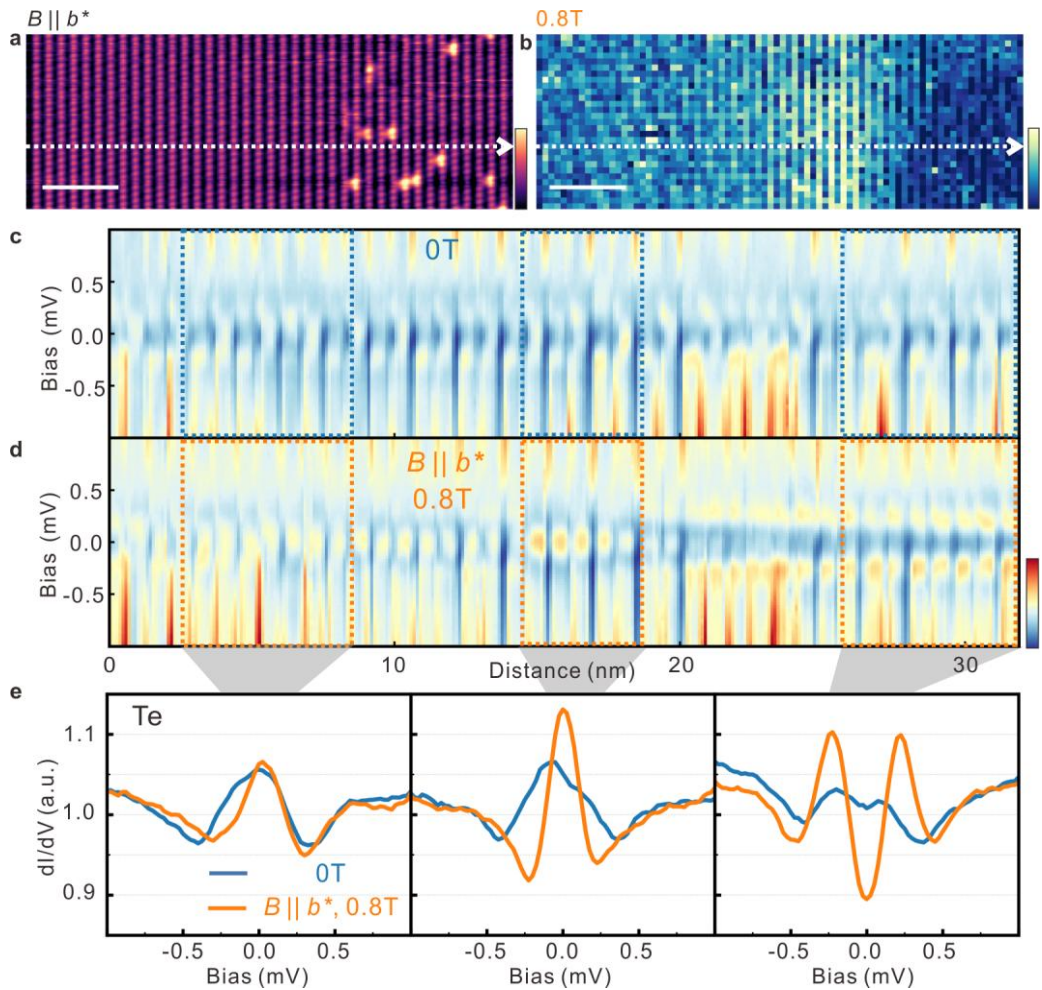

**Supplementary Figure 8 | Out-of-plane vortex on the UTe$_2$ surface under magnetic field along the *b*\*-axis. a,** Topography of the vortex region ($V$ = 40 mV, $I$ =20 pA). Scale bar, 5 nm. **b,** Corresponding d$I$/d$V$ map acquired at 0 mV under a 0.8 T magnetic field applied along the *b*\*-axis. An asymmetric vortex appears and is pinned by nearby defects ($V$ = 1 mV, $I$ =100 pA). Scale bar, 5 nm. **c,** d$I$/d$V$ line cut measured along the dashed line in **a** under zero field ($V$ = 1 mV, $I$ =100 pA). **d,** d$I$/d$V$ line cut measured along the dashed line in **b**, which corresponds to the real-space dashed line shown in **a**, under a 0.8 T magnetic field applied along the *b*\* axis ($V$ = 1 mV, $I$ =100 pA). A zero-energy vortex bound state appears in the center. **e,** Averaged spectra of Te sites extracted from the d$I$/d$V$ linecuts corresponding to the regions marked by dashed boxes in **c** and **d** under the indicated fields. The gap deepens on the right side of the vortex under field, remains largely unchanged on the left, and a sharp zero-energy peak emerges at the vortex core.

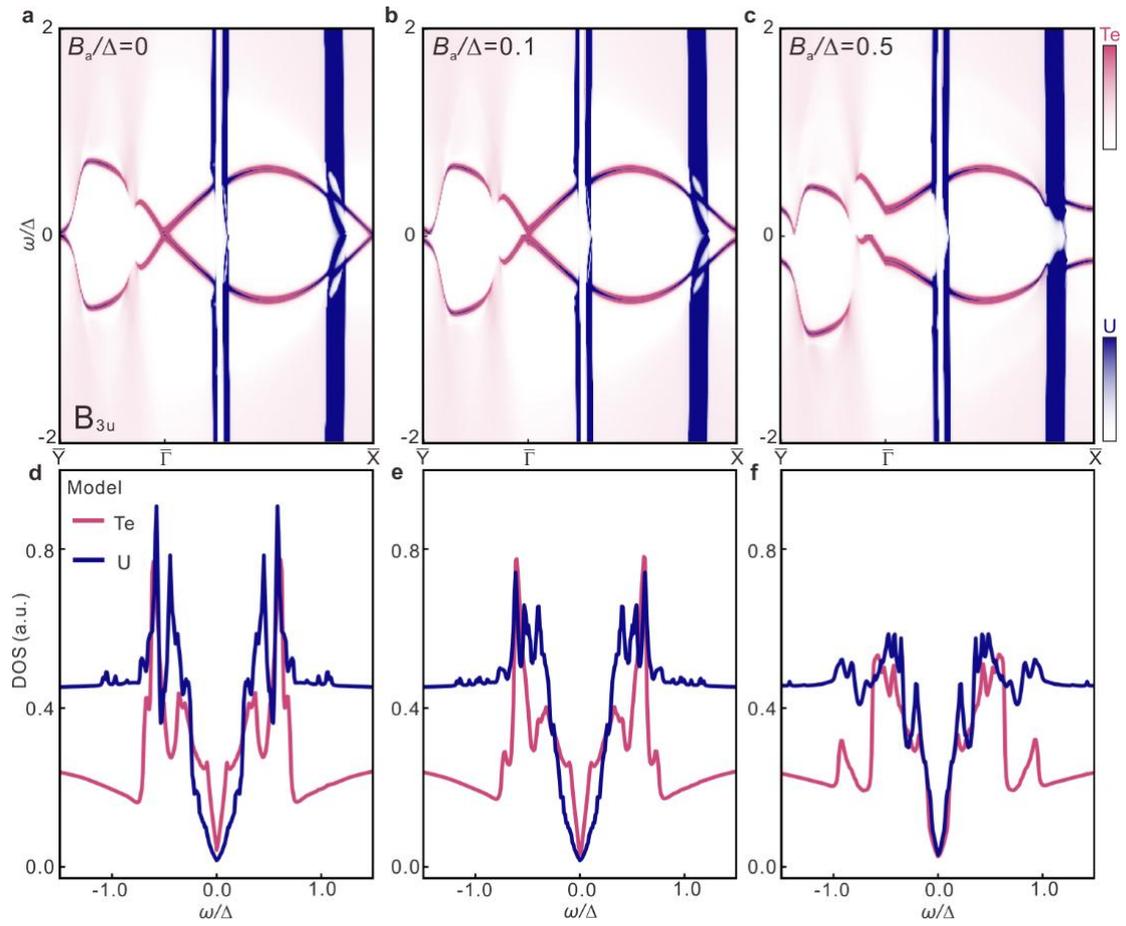

**Supplementary Figure 9 | Zeeman effect on the topological surface states of UTe₂ in the $B_{3u}$ triplet phase. a-c,** Spectral functions of Te and U in the $B_{3u}$ triplet phase with the indicated Zeeman energy. The topological surface states dominated by Te are gradually gapped out by the Zeeman effect. **d-f,** Integrated DOS versus energy corresponding to **a-c**. As Zeeman energy increases, the Te-derived sub-gap DOS associated with the topological surface states becomes suppressed.